\documentclass[preprint,pre,showpacs,epsfig,address]{revtex4}%
\usepackage[english]{babel}
\usepackage[latin1]{inputenc}
\usepackage[dvips]{graphicx}
\usepackage{amsmath}
\usepackage{amssymb}
\usepackage{epsfig}
\usepackage{array}
\usepackage{amsfonts}
\usepackage{graphicx}%
\begin{document}
\title{Hydrodynamic limit of multi-chain driven diffusive models}
\author{V. Popkov}
\affiliation{Institut f\"ur Festk\"orperforschung, Forschungszentrum J\"ulich - 52425
J\"ulich, Germany}
\author{M. Salerno }
\affiliation{Dipartimento di Fisica ``E.R. Caianiello" and Istituto Nazionale di Fisica
della Materia (INFM), Universit\'a di Salerno, I-84081 Baronissi (SA), Italy}
\date{\today }

\begin{abstract}
A new class of models, generalizing Asymmetric Exclusion Process for many
parallel interacting channels, is proposed. We couple the models with boundary
reservoirs, study boundary-driven phase transitions and show that usually
taken hydrodynamic description fails. The adequate hydrodynamic limit is then
derived. We support our findings with Monte-Carlo simulations of the original
stochastic system.

\end{abstract}
\pacs{05.70.Ln, 64.60.Cn, 02.50.Ga }

\maketitle

\section{Introduction}

Many interesting non-equilibrium physical phenomena (shock wave dynamics,
boundary driven phase transitions, etc.) can be observed already in simplest
models such as driven diffusing particle systems, recent review of which can
be found e.g. in \cite{Schu00,Liggett1999}. These systems have a unique
ability to feel the dynamics of the boundaries thanks to the presence of a
flux of particles which brings information from boundaries to the bulk. In
absence of a flux, indeed, boundaries would play marginal role, as in ordinary
equilibrium statistical mechanics, but in presence of a flux the boundaries
may dominate the bulk with the possibility to give rise to boundary driven
phase transitions. These phenomena were indeed observed in one-species driven
diffusive models \cite{Krug91,Kolo98, Gunter_Slava_Europhys} and in some
models containing more than one species of particles (many-species models)
\cite{Mukamel95, Peschel}. In a more general context, the problem was
addressed in \cite{Kolo98}, where it was shown that the stationary state of
systems with one species of particles with open boundaries in non-equilibrium
depends only on the stationary flux. This implies that an extremum principle
for the stationary flux can be formulated \cite{Gunter_Slava_Europhys} and the
properties of the stationary state can be obtained directly from the
hydrodynamic limit. In this approach the stochastic model is mapped into a
viscousless conservation law equation which is ill-defined (the corresponding
Cauchy problem admits multiple solutions). Since the dynamics of stochastic
system (Markov process) is unique, the problem of regularization, i.e. how to
single out the physical solution from a set of multiple solutions, arises. As
is well known, this problem is resolved by adding a phenomenological small
dissipative term (of order $\varepsilon$) to the conservation law equation and
then taking the limit $\varepsilon\rightarrow0$. For this purpose, a linear
viscous term of the form $\varepsilon$ $\partial^{2}u/\partial x^{2}$ is
usually considered, although nonlinear terms of the type $\varepsilon$ $f(u)$
$\partial^{2}u/\partial x^{2}$ , with $f(u)$ being an arbitrary convex
function, are also possible. For one-species systems this regularization
procedure is quite robust in the sense that different choices of the viscosity
term can be shown to lead to the same physical solution. For many-species
models, however, this is not trivial. In this case the hydrodynamic limit is
is described by a system of conservation laws (one for each particle species)
and the regularization is usually achieved by adding a diagonal diffusion
matrix of the form $\varepsilon$ $\partial^{2}u^{i}/\partial x^{2}$, to the
system. In spite of its simplicity, there is no evidence that this approach
works also for these more complicated systems and the regularization problem
of multi-species driven diffusive systems is still open.

The aim of the present Letter is two-fold. From one side, we introduce a new
class of multi-species driven diffusive models which generalize the single
chain asymmetric exclusion process (ASEP). These models possess nice
properties such as product measure stationary state and particle-hole
symmetry. From a physical point view, they describe the motion of particles in
multi-channel cables with the particles in adjacent cables creating an
effective barrier potential for the particles in a given cable to flow. From
the other side, we study the hydrodynamic limit of these models by showing
that the conventional regularization of the conservation law equations leads
to wrong hydrodynamic results. General arguments which support this failure
will be provided. On the contrary, we show that an alternative regularization,
obtained directly from the microscopic dynamics, leads to correct results. A
detailed comparison of the hydrodynamic predictions (both with conventional
and alternative regularization) with direct Monte Carlo simulation of the
original stochastic model confirms these results.

Although the mapping from \textit{hydrodynamic equations} to
\textit{\ stochastic processes} is not unique, it is likely that our
alternative regularization may work also for other multi-species driven
diffusive models.

\section{The model}

\begin{figure}[ptb]
\begin{center}
\includegraphics[scale=0.9]{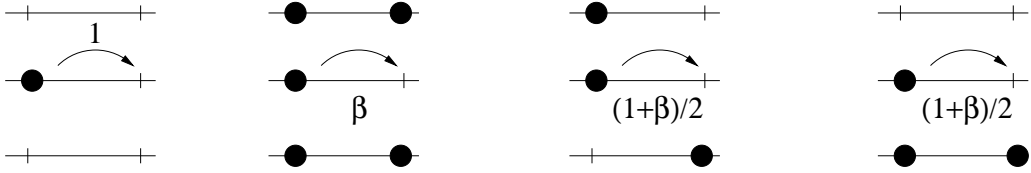}
\end{center}
\caption{Elementary hopping processes happening in the case of $M=3,S=2$, with
their rates. Four out of $16$ possible configurations are shown. The rates
(\ref{rates}) depend only on the total number of particles in the neighboring
four sites.}%
\label{fig_16_rates}%
\end{figure}

To introduce the model, we consider $M\geq2$ discrete chains on which
particles can hop preferentially in one direction. The hopping between
adjacent chains is forbidden but particles can move from a site $k$ to an
empty site $k+1$ on the same chain with rates that depend on the particle
configuration at adjacent sites of "neighboring" $S$ chains (the results will
be qualitatively the same for other possibilities of next-neighboring
hopping). Notice that such a dynamics does not satisfy detailed balance
condition meaning that the system is far from thermodynamic equilibrium. The
hopping rates for different configurations can be expressed in terms of a
single parameter $\beta$ measuring the strength of the interaction between the
chains. Let us denote with $r_{n}$ the hopping rate from site $k$ to site
$k+1$ on the same chain, in presence of a total number $n$ of particles in
adjacent chain sites. We restrict our search to the models having stationary
product measures. In this case, calculation shows that the rates $r_{n}$ have
to satisfy the condition $r_{n}-r_{n-1}=const$ for any $n$. Moreover, the
rates $r_{n}$ turn out to depend only on $n$ and not on the particular
configurations the $n$ particles can assume. With $S$ chains-neighbors, any
given pair of consecutive sites $k,k+1$ can have from $n=0$ up to $n=2S$
particles-neighbors. An example of the hopping process with the corresponding
rates is shown in Fig.~\ref{fig_16_rates} for the case $S=2$. In general, for
a chain having $S$ chains-neighbors \cite{differentS}, the rates of hopping
(normalized to $r_{0}=1$) can be parametrized as
\begin{equation}
r_{n}=1+n(\beta-1)/2S,\text{ \ \ \ \ \ \ \ }n=0,1,...,\,2S.\label{rates}%
\end{equation}
Notice that for $\beta=1$ the rates become independent on $n$ (i.e. the
interchain interaction becomes zero) and the system splits into $M$ parallel
uncoupled totally asymmetric exclusion processes (TASEP)
\cite{ASEP,Liggett1999}. For $\beta<1$, the rates (\ref{rates}) monotonically
decrease with number of particles-neighbors $n$, i.e. the presence of adjacent
particles creates an effective barrier potential which slows down the particle movement.

The driven diffusive models described by (\ref{rates}) possess the
following properties. \textit{i) Product measure.} One can easily check that
the model admits a stationary distribution in the class of product measures.
This means that the stationary state of the model is spatially uncorrelated,
both in the longitudinal and in the transverse direction. This property allows
to obtain a simple analytic expression for the stationary flux (see below).

\textit{ii) Particle - hole symmetry.} It is easy to check that an exchange of
particles with holes plus the substitution $\beta\rightarrow1/\beta$ leaves
the model invariant. Since parameter $\beta$ must be is in the range
$0\leq\beta<\infty$ (the rates in (\ref{rates}) are nonnegative), the present
property allows to further restrict $\beta$ to the range $0\leq\beta\leq1$
only. In the limit of the strongest interaction $\beta=0$ the hopping of a
particle in completely saturated environment becomes impossible $r_{2S}=0$.
Let us consider some specific examples.

\noindent\textit{Case $S=1$.} In this case there is one "neighboring" chain
for each given chain. This can happen if there are only two chains ($M=2$) or
if in the model with $M$ consecutive chains, the hopping along one chain
depends only on the state of the next chain. Due to the property (i), the
stationary flux on a given chain for $S=1$ can be computed straightforwardly
as
\begin{equation}%
\begin{array}
[c]{cc}%
j(\rho,\sigma) & =\rho(1-\rho)(1+(\beta-1)\sigma)
\end{array}
\label{fluxS1}%
\end{equation}
where $\rho$ and $\sigma$ are average densities of particles on a given chain
and on a chain "neighboring" to it, respectively. The case $M=2$ was
considered in \cite{GunterJSP} where the properties of elementary excitations
were studied.

\noindent\textit{Case $S=2$.} In this case each pair of sites on two
neighboring chains have up to $4$ particle-neighbors (see
Fig.~\ref{fig_16_rates}). this situation can be realized such a in the
geometry of $M $ consecutive chains with periodic boundary conditions
$M+1\equiv1$ \ in the transverse direction, where each chain $m$ has two
chains-neighbors $m\pm1$. The hopping rates are then obtained as
\begin{equation}
r_{n}=\left\{
\begin{array}
[c]{c}%
1\text{ \ \ \ \ \ \ \ \ \ \ \ \ \ \ \ if \ }n=0,\\
(\beta+3)/4\text{ \ \ \ if \ }n=1,\\
(\beta+1)/2\text{\ \ \ \ if \ }n=2,\\
(3\beta+1)/4\text{ \ if }n=3,\\
\beta\text{ \ \ \ \ \ \ \ \ \ \ \ \ \ \ \ if }n=4.
\end{array}
\right.  \label{rates16}%
\end{equation}
The stationary flux on a chain $m$ in a system with average density $\rho_{k}
$ on the chain $k$ can be obtained using property (i), as%
\begin{equation}
j_{m}=\rho_{m}(1-\rho_{m})(1+\frac{1}{2}(\beta-1)(\rho_{m+1}+\rho_{m-1}))
\label{fluxS2}%
\end{equation}

\noindent\textit{Case $S>2$.} For arbitrary number of $S$ chains-neighbors to
a given one, the stationary flux is given by
\begin{equation}
j_{m}=\rho_{m}(1-\rho_{m})\left(  1+\frac{1}{S}(\beta-1){\displaystyle\sum
\limits_{neighbors}}\rho_{k}\right)  , \label{fluxS}%
\end{equation}
where the sum is taken over the "neighboring" chains to a chain $m$, and
$\rho_{k}$ is an average particle density on chain $k$.

From a physical point of view the case $S>2$ corresponds to a coaxial cable
with many fibers, on which particles move, each fiber having $S$
fibers-neighbors surrounding it. A real system is always finite, so that
boundary conditions, where particles can enter or exit the system must be
imposed. To model boundaries, we couple model (\ref{rates}) with stationary
particle reservoirs of densities $\rho_{L}^{m}$ on the left boundary of chain
$m$, from where particles can be injected , and of the densities $\rho_{R}%
^{m},$ on the right boundary of it, where they can be extracted. The rates of
extraction and injection are obtained from the boundary densities and for
specific cases were given in \cite{Peschel,Hager} .

To proceed further, we fix the total number of chains, $M$, the number of
chains-neighbors of a given chain $S$, the right and left boundary densities
$\rho_{R}^{m},\rho_{L}^{m},$ and study bulk stationary densities that is, in
the limit when time tends to infinity $t\rightarrow\infty$ \cite{Peschel_case}%
, as the interchain interaction varies from $\beta=1$ (non interacting case)
to $\beta=0$ (case of maximal interaction).

\begin{figure}[ptb]
\begin{center}
\includegraphics[scale=0.7]{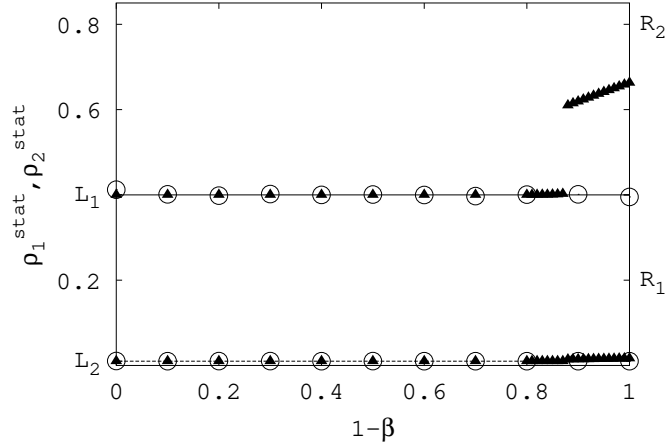}
\end{center}
\caption{ Comparison of predictions of MonteCarlo (open circles), the
hydrodynamic equations Eq. (\ref{hydro})(line) and Eq.(\ref{cons2})(filled
triangles) for stationary densities of particles in the system as a function
of interchain interaction $\beta-1$. Parameters are: $M=2,S=1$, the boundary
densities of the chains are $\rho_{L}^{1},\rho_{L}^{2}=0.4,0.01$ on the left
and $\rho_{R}^{1},\rho_{R}^{2}=0.2,0.8$ on the right end.}%
\label{fig_piccolo_chain2}%
\end{figure}

It is known that for an Asymmetric Exclusion process, which is a limit of our
models for $\beta=1$, \ there is a first order non-equilibrium phase
transition at $\rho_{R}^{m}=1-\rho_{L}^{m}>\frac{1}{2}$, where discontinuous
transition from the low density phase $\rho_{stat}^{m}=\rho_{L}^{m}$ for
$\rho_{R}^{m}=(1-\rho_{L}^{m})-0$ to a high density phase $\rho_{stat}%
^{m}=\rho_{R}^{m}$ for $\rho_{R}^{m}=(1-\rho_{L}^{m})+0$ happens. We will
choose one of the boundary densities $\rho_{R}^{m},\rho_{L}^{m}$ in the
vicinity of this phase transition point (for $\beta=1$), different for
different $m$ \ to exclude possible degeneracies, and look at the dependence
of $\rho_{stat}^{m}(\beta)$ for different $M,S$. Alternatively, we shall look
at the dynamic properties of the stochastic model and the hydrodynamic
equation, comparing the time evolution of an initial state after certain time.
The results of the Monte Carlo calculations can be seen on the graphs
Figs.\ref{fig_piccolo_chain2},\ref{fig_piccolo_chain3} alongside with the
results of numerical integration of corresponding partial differential
equations arising from stochastic model in hydrodynamic limit. Derivation of
these equations is given below.

\section{Hydrodynamic limit}

\begin{figure}[ptb]
\begin{center}
\includegraphics[scale=0.7]{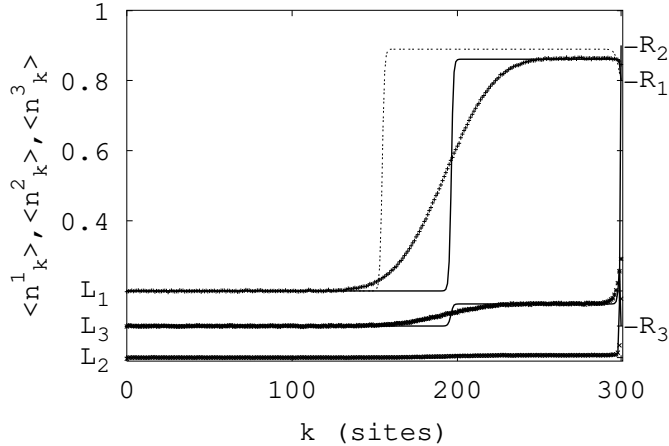}
\end{center}
\caption{ The time evolution of the stochastic model (\ref{rates}) with
$\beta=0$, $M=3$. Comparison of average density profiles given by Monte-Carlo
simulation (points), hydrodynamic equations (\ref{hydro}) (lines), and naive
viscosity approach (\ref{cons2})(broken line). The system has evolved for
$t=1600$ Monte Carlo steps from homogeneous initial condition $\langle
n_{k}^{Z}\rangle=\rho_{L}^{Z}$, matching the left boundary densities $\rho
_{L}^{1}=0.2,\rho_{L}^{2}=0.01,\rho_{L}^{3}=0.1$. The corresponding right
boundary particle densities are $0.8/0.9/0.1$. An average over $10^{5}$
histories is taken. One sees that neither the position nor the level of the
shock are predicted correctly by the naive viscosity approach (the broken
line).}%
\label{fig_piccolo_chain3}%
\end{figure}

The naive continuum (Eulerian) limit of our stochastic dynamics on the lattice
is a system of conservation laws
\begin{equation}
{\frac{\partial\rho^{Z}(x,t)}{\partial t}}+{\frac{\partial j^{Z}}{\partial x}%
}=\epsilon{\frac{\partial^{2}\rho^{Z}}{\partial x^{2}}};\ \ \ Z=1,2,...,\,M,
\label{cons2}%
\end{equation}
with a phenomenological vanishing viscosity $\varepsilon\rightarrow0$
(regularization term) on the right hand side added. Here $\rho^{Z}(x,t)$
denote coarse-grained particle densities while $\ j^{Z}$ --- the particle
fluxes given by Eq. (\ref{fluxS}). The regularization term in Eq.
(\ref{cons2}) leads to the correct answer for the initial Riemann problem, as
compared to the stochastic model (see \cite{GunterJSP}) but fails to describe
the reflection from the boundaries (see \cite{VP_in_preparation},
Figs.\ref{fig_piccolo_chain2},.\ref{fig_piccolo_chain3}).

An adequate regularization is obtained by averaging exact lattice continuity
equations of the stochastic process
\begin{equation}
{\frac{\partial}{\partial t}}\hat{n}_{k}=\hat{\jmath}_{k-1}-\hat{\jmath}_{k}
\label{lattice_continuity_equation}%
\end{equation}
for occupation number operators $\langle\hat{n}_{k}\rangle\rightarrow
\rho(x,t)$ of any given chain, making the lattice constant infinitesimally
small $k,k+1\rightarrow x,\ x+a$. For the case (\ref{rates}), the local flux
operator $\hat{\jmath}_{k}$ between the sites $k$ and $k+1$ reads (see
\cite{GunterJSP} for a derivation of (\ref{flux_operator}) in the case
$M=2,S=1$):
\begin{equation}
\hat{\jmath}_{k}=\hat{n}_{k}(1-\hat{n}_{k+1})\left(  1+{\frac{\beta-1}{2S}%
}{\displaystyle\sum\limits_{p=1}^{S}}\left(  \hat{m}_{k}^{p}+\hat{m}_{k+1}%
^{p}\right)  \right)  . \label{flux_operator}%
\end{equation}
Here $\hat{m}_{k}^{p},$ \ $p=1,...S$ are occupation number operators for a
site $k$ of the chains neighboring to a given one. We substitute
(\ref{flux_operator}) into (\ref{lattice_continuity_equation}), average,
factorize and Taylor expand the latter with respect to the lattice constant
$a$ according to $\langle\hat{n}_{k+1}\rangle=\rho(x,t)+a\frac{\partial\rho
}{\partial x}+\frac{a^{2}}{2}\frac{\partial^{2}\rho}{\partial x^{2}}+O(a^{3}%
)$, etc. Expanding the resulting equation in powers of $a$ and keeping terms
up to the second order, we obtain the following hydrodynamic equation
\begin{equation}
{\frac{\partial\rho^{Z}(x,t)}{\partial t}}+{\frac{\partial j^{Z}}{\partial x}%
}=\epsilon{\frac{\partial}{\partial x}}\left(  \frac{\partial\rho^{Z}%
}{\partial x}\left(  1+{\frac{\beta-1}{S}}{\displaystyle\sum\limits_{p=1}^{S}%
}\rho^{p}\right)  \right)  ;\ \ \ Z=1,2,...M, \label{hydro}%
\end{equation}
where $\rho^{p}=\rho^{p}(x,t)$ are local densities of the chains neighboring
to a given chain $Z,$ $\varepsilon={\frac{a}{2}}\rightarrow0;\ \ {\frac
{\partial}{\partial t}}\rightarrow2\varepsilon{\frac{\partial}{\partial t}}$.
Notice that Eq. (\ref{hydro}) is valid for arbitrary $M,S$ and for $M=2,S=1$
coincides with the hydrodynamic equation derived in Ref. \cite{GunterJSP}.

To check the validity of this approach we have compared direct numerical
integrations of (\ref{hydro}) with Monte Carlo simulations of the original
stochastic model described by the rates in Eq. (\ref{rates}). As a result we
obtain that, while the agreement between the Monte Carlo predictions and the
numerical integration of Eq. (\ref{hydro}) is excellent, the comparison with
the standard hydrodynamic limit gives rise to inconsistencies (see discussion
below) both in the stationary regime and during relaxation towards
stationarity. This is shown in Fig.\ref{fig_piccolo_chain2} for system of two
and in Fig.\ref{fig_piccolo_chain3} for three chains with $S=1$. From these
figures it is clear that the choice of the viscosity is indeed crucial to
obtain correct results, since different choices will produce different
answers. We checked that Eq. (\ref{hydro}) gives results which coincide, up to
non-universal boundary layer and interface width, with Monte Carlo simulations
also for other values of $M,S$, and we believe this fact is valid in general.

The above analysis can be easily generalized to the case in which the hopping
along the chains occurs in both directions. In this case the corresponding
flux to the left is obtained from the one to the right by exchanging
$\rho(1-\rho)$ in (\ref{fluxS}) with the term $(p-q)\rho(1-\rho)$, where $p$
and $q$ denote the particle hopping rates on the $m$-th chain to the right and
to the left in the empty environment (the general conditions for a product
measure state can be found in \cite{Toth}).

It is interesting to discuss why a diagonal diffusion matrix leads to
inconsistent results for multi-species systems. To this regard, we remark that
one can formally obtain a diagonal diffusion matrix in Taylor expansion of
(\ref{lattice_continuity_equation}) by adding to the flux (\ref{flux_operator}%
) an additional term as, e.g.%
\begin{equation}
\hat{\jmath}_{k}=\hat{n}_{k}(1-\hat{n}_{k+1})\left(  1+{\frac{\beta-1}{2S}%
}{\displaystyle\sum\limits_{p=1}^{S}}\left(  \hat{m}_{k}^{p}+\hat{m}_{k+1}%
^{p}\right)  \right)  +\left(  \frac{\beta-1}{4S}\left(  \hat{n}_{k+1}-\hat
{n}_{k}\right)  {\displaystyle\sum\limits_{p=1}^{S}}\left(  \hat{m}_{k}%
^{p}+\hat{m}_{k+1}^{p}\right)  \right)  . \label{naive}%
\end{equation}
If one substitutes the above expression into
(\ref{lattice_continuity_equation}) and performs the same analysis as before,
one obtains the resulting equation in the form (\ref{cons2}). In this case,
however, the factorization becomes invalid because the flux operator
(\ref{naive}) describes different process with correlations in the stationary
state and therefore with a stationary flux different from (\ref{fluxS}). Thus,
by making the diffusion matrix diagonal one expects inconsistencies. This can
be seen by considering a microscopic state with all but one chains completely
filled with particles, the chain partially filled having an empty region
separated by a completely filled one, $\hat{n}_{k}=0$, \ $k\leq0$ and $\hat
{n}_{k}=1$, \ $k>0$. According to the dynamic rules in Fig.\ref{fig_16_rates},
no movement is allowed, while from Eq. (\ref{lattice_continuity_equation})
(with averaged and factorized flux as in Eq. (\ref{naive})), we obtain that
${\frac{\partial}{\partial t}}\hat{n}_{0}\neq0$. Notice that although the
choice of the second term in Eq. (\ref{naive}) is not unique, the above
arguments for inconsistency would still be valid in presence of alternative
choices. These inconsistencies are clearly seen on
Fig.\ref{fig_piccolo_chain2}, where a comparison between stationary densities
of the two chain system ($M=2)$ obtained from Monte Carlo calculations and
hydrodynamic equations (\ref{cons2},\ref{hydro}) is made, as a function of the
interchain interaction $1-\beta$. The naive description (\ref{cons2}) shows a
phase transition at $\beta\approx0.1$ , while Monte Carlo and \ (\ref{hydro})
show no effect. The occurrence of a phase transition to high densities can be
understood as follows. Since the right boundary density of particles of chain
$2$ is high ($\left(  \rho_{2}\right)  _{R}=0.8$, it becomes more and more
difficult for the particles of chain $1$ to exit the chain with increasing
interaction, so that "traffic jam" occurs at the exit, this leading to the
phase transition. This effect is greatly overestimated by the naive
hydrodynamic description ( \ref{cons2}) (in physical system the transition
happens for higher left boundary density $\left(  \rho_{1}\right)  _{L}%
\gtrsim0.48$).

%In panel b of Fig.\ref{fig_piccolo_chain2} we compare the average
%density profiles computed with the different methods, for another
%model of the same class (\ref{rates}). Again, one sees that the
%prediction given by the naive viscosity approach (\ref{cons2})
%leads to wrong results.

\section{Conclusion}

In conclusion, we have introduced a class of models which generalize the
asymmetric simple exclusion process for an arbitrary number of chains and with
hopping rates given as a function of the local configuration in the
neighboring chains (\ref{rates}) describing the effective friction a particle
encounters moving in a dense environment. For these models we derived the
stationary flux, and modified the conventional choice of the viscosity term in
the hydrodynamic limit (see, e.g.,\cite{Bressan}) in order to get a good
agreement with Monte Carlo simulations of the original stochastic process.
Although the problem of the hydrodynamic limit has been discussed on the
specific example of the introduced models, the results are expected to be
valid in general (the advantage of using our model is only for computational
convenience since the product measure property allows to obtain the bulk flux,
boundary conditions, etc., in an easy and straightforward manner.)

\begin{acknowledgments}
V.P. thanks G.M. Sch\"{u}tz for fruitful discussions, and the Department of
Physics at the University of Salerno for hospitality. Financial support from
the Deutsche Forschungsgemeinschaft, and from the three month grant of the
Istituto Nazionale di Fisica della Materia (INFM) unit\'{a} di Salerno is
gratefully acknowledged
\end{acknowledgments}


\begin{thebibliography}{99}                                                                                               %


\bibitem {Schu00}G.M. Sch\"{u}tz \emph{Exactly solvable models for many-body
systems far from equilibrium}, in: \emph{Phase Transitions and Critical
Phenomena} , C. Domb and J. Lebowitz (eds) (Academic, London, 2000 ) vol.
\textbf{19.}

\bibitem {Liggett1999}T. M. Liggett \textit{Stochastic interacting systems:
contact, voter and exclusion processes} (Springer, Grundlehren der
Mathematischen Wissenschaften, 1999) vol. \textbf{324.}

\bibitem {Krug91}J. Krug \textit{Phys. Rev. Lett.} \textbf{67,} 1882 (1991).

\bibitem {Kolo98}A. B. Kolomeisky , G. M. Sch\"{u}tz , E. B. Kolomeisky and J.
PStraley \textit{J. Phys. A} \textbf{31,} 6911 (1998).

\bibitem {Gunter_Slava_Europhys}V. Popkov and G. M. Sch\"{u}tz
\textit{Europhys. Lett.} \textbf{48,} 257 (1999).

\bibitem {Mukamel95}M. R. Evans , D. P. Foster , C. Godr\`{e}che and D.
Mukamel \textit{Phys. Rev. Lett.} \textbf{74,} 208 (1995) --- \textit{J. Stat.
Phys.} \textbf{80,} 69 (1995).

\bibitem {Peschel}V. Popkov and I. Peschel \textit{Phys. Rev. E } \textbf{64,}
026126 (2001).

\bibitem {differentS}Different chains may have different number of
chains-neighbors (e.g., due to topological reasons). In this case, in order to
have a product stationary state in the system, one should require
$r_{n}-r_{n+1}=const$ for the rates of hopping along all the chains. Clearly,
this condition is satisfied by (\ref{rates})

\bibitem {ASEP}G. Sch\"{u}tz and E. Domany \textit{J. Stat. Phys.}
\textbf{72,} 277 (1993); B. Derrida , M. R. Evans , V. Hakim and V. Pasquier
\textit{J.Phys.A} \textbf{26,} 1493 (1993).

\bibitem {GunterJSP}V. Popkov and G. M. Sch\"{u}tz \textit{J Stat. Phys
}\textbf{112}, 523 (2003).

\bibitem {Hager}J. S. Hager , J. Krug , V. Popkov and G. M. Sch\"{u}tz ,
\textit{Phys. Rev. E} \textbf{63}, 056110 (2001).

\bibitem {Peschel_case}Note that there are cases when stationary state has
complicated structure, e.g., spatial coexistence of structurally different
regions \cite{Peschel}. However in the models (\ref{rates}) that we consider,
bulk density is a good order parameter for a stationary state description.

\bibitem {VP_in_preparation}V. Popkov, in preparation

\bibitem {Toth}B. T\'{o}th and B. Valk\'{o} \textit{J Stat. Phys }%
\textbf{112}, 497 (2003).

\bibitem {Bressan}A. Bressan, \textit{Hyperbolic systems of conservation
laws}, Oxford University Press, 2000).
\end{thebibliography}
\end{document}